
\input harvmac
\noblackbox
%
%

\def\half{{1\over2}}

\def\apm{\alpha^{\prime}}

\def\p{\partial}

%
%
\lref\serg{S. Cecotti, S. Ferrara and L. Girardello,
``Geometry of Type II Superstrings and the Moduli of
Superconformal Field Theories'', Int. J. Mod. Phys. {\bf 4} (1989) 2475.}
\lref\bwit{J. Bagger and E. Witten, Nucl. Phys. {\bf B222} (1983) 1.}
\lref\master{P. Breitenlohner, D. Mason and G. Gibbons, ``4-Dimensional
Black Holes from Kaluza-Klein Theories'', Comm. Math. Phys. {\bf 120}
(1988) 295.}
\lref\ascmp{A. Strominger, ``Special Geometry'', Comm. Math. Phys. {\bf 133}
(1990) 163.}
\lref\shenker{S. Shenker, ``The Strength of Nonperturbative Effects in String
Theory'',
in proceedings of the Cargese Workshop on Random Surfaces, Quantum Gravity
and Strings (1990).  }
\lref\bbs{K. Becker, M. Becker and A. Strominger, unpublished.}
\lref\gihu{G.W. Gibbons and C.M. Hull, Phys. Lett. {\bf 109B} (1982) 190.}
\lref\dlwp{B. de Wit, P. Lauwers and A Van Proeyen, ``Lagrangians of N=2
Supergravity-Matter Systems'', Nucl. Phys. {\bf B255} (1985) 269.}
\lref\conif{P. Candelas, P. Green and T. Hubsch, ``Rolling
Among Calabi-Yau Vacua'', Nucl. Phys. {\bf B330} (1990) 49.   }
\lref\tian{G. Tian, in {\sl Mathematical Aspects of String Theory}, ed.
S. T. Yau, World Scientific (1987) 629.}
\lref\phxe{P. Candelas and X. de la Ossa, ``Moduli Space of Calabi-Yau
Manifolds'', Nucl. Phys. {\bf B42 } (1990) 246.}
\lref\phxec{P. Candelas and X. de la Ossa, ``Comments on Conifolds'',
Nucl. Phys. {\bf B42 } (1990) 246.}
\lref\DGHR{ A. Dabholkar, G. Gibbons, J. Harvey and F. R. Ruiz,
{\it Superstrings and Solitons}, Nucl. Phys. {\bf B340} (1990) 33.}
\lref\hetsol{ A. Strominger, {\it Heterotic Solitons},
Nucl. Phys. {\bf B343} (1990) 167; E: Nucl. Phys. {\bf B353} (1991) 565.}
\lref\jhasl{ J. Harvey and A. Strominger, ``The
Heterotic String is a Soliton'', hep-th/9504047.}
\lref\town{C.M. Hull and P.K.  Townsend, ``Unity of Superstring
Dualites'', QMW-94-30, R/94/33,
hep-th/9410167. }
\lref\dhs{M. Dine, P. Huet, and N. Seiberg,
``Large and Small Radius in String Theory'',
Nucl. Phys. {\bf B322} (1989) 2073. }
\lref\duffone{M. J. Duff and J. X. Lu,
{\it Elementary Fivebrane Solutions of D=10
Supergravity}, Nucl. Phys. {\bf B354} (1991) 141.}
\lref\dufftwo{M. J Duff and J. X. Lu, {\it Remarks on String/Fivebrane
Duality}, Nucl. Phys. {\bf B354} (1991) 129.}
\lref\duffthree{M. J. Duff and J. X. Lu, {\it Strings from Fivebranes},
Phys. Rev. Lett. {\bf 66} (1991) 1402.}
\lref\duffthrb{M. J. Duff and J. X. Lu, ``The Self-Dual Type IIB
Superthreebrane'',
Phys. Lett. {\bf 273} (1991) 409.}
\lref\wit{E. Witten, ``String Theory Dynamics in Various Dimensions'',
hep-th/9503124.}
\lref\duff{M. J. Duff and J. X. Lu, Nucl. Phys. {\bf B357} (1991) 534.}
\lref\duffss{M. J. Duff, ``Strong/Weak Coupling Duality from
the Dual String'', NI-94-033,
CTP-TAMU-49/94, hep-th/9501030.}
\lref\SCH{J. H. Schwarz, Nucl. Phys. {\bf B226} (1983) 269. }
\lref\drev{For a review see M. J. Duff,  R. R. Khuri, and J. X. Lu, ``String
Solitons,'' hep-th/9412184.}
\lref\sentwo{A. Sen, ``String-String Duality Conjecture in Six Dimensions and
Charged Solitonic Strings'', hep-th/9504027.}
\lref\nati{N. Seiberg, ``Observations on the Moduli Space of Superconformal
Field Theories'', Nucl. Phys. {\bf B303}, (1986), 288;  P. Aspinwall and D.
Morrison,  ``String
Theory on K3 Surfaces'',  preprint DUK-TH-94-68, IASSNS-HEP-94/23,
hep-th/9404151.}
\lref\swit{N. Seiberg and E. Witten, ``Electromagnetic Duality,
Monopole Condensation and Confinement in N=2 Supersymmetric
Yang-Mills Theory'', hep-th/9407087, Nucl. Phys. {\bf B426}, (1994), 19.}
\lref\ferr{A. Ceresole, R. D'Auria, S. Ferrara and A. Van Proeyen,
``Duality Transformations in Supersymmetric Yang-Mills Theory Coupled
to Supergravity, '' hep-th/9502072.}
\lref\DAHA{ A. Dabholkar and J. A. Harvey,  ``Nonrenormalization of
the Superstring Tension, '' Phys. Rev. Lett.
{\bf 63} (1989) 719.}
\lref\TEN{ I. C. G. Campbell and P. West, Nucl. Phys. {\bf B243},
(1984) 112; F. Giani and M. Pernici, Phys. Rev. {\bf D30}, (1984)
325;
M. Huq and M. A. Namazie, Class. Quant. Grav. {\bf 2} (1985) 293.}
\lref\DGHR{ A. Dabholkar, G. Gibbons, J. A. Harvey and F. R. Ruiz,
``Superstrings and Solitons, '' Nucl. Phys. {\bf B340} (1990) 33.}
\lref\duffive{ M. J Duff,  ``Supermembranes, the First Fifteen Weeks, ''
Class. Quant. Grav. {\bf 5} (1988) 189.}
\lref\hetsol{ A. Strominger,  ``Heterotic Solitons, ''
Nucl. Phys. {\bf B343} (1990) 167; E: Nucl. Phys. {\bf B353} (1991) 565.}
\lref\bhole{G. Horowitz and A. Strominger, ``Black Strings and
$p$-branes,'' Nucl. Phys. {\bf B360} (1991) 197.}
\lref\wbran{C. G. Callan, J. A. Harvey and A. Strominger,
``Worldbrane Actions for String Solitons,'' Nucl. Phys. {\bf B367} (1991) 60.}
\lref\tei{R. Nepomechie, Phys. Rev. {\bf D31} (1985) 1921;
C. Teitelboim, Phys. Lett. {\bf B176} (1986) 69.}
\lref\wsheet{C.G. Callan, J. A. Harvey and A. Strominger,
``Worldsheet Approach to Heterotic Instantons and  Solitons,'' Nucl. Phys. {\bf
359}
(1991) 611. }
\lref\duffcq{M. J. Duff and J. X. Lu,  ``Remarks on String/Fivebrane
Duality,'' Nucl. Phys. {\bf B354} (1991) 129; ``String/Fivebrane Duality, Loop
Expansions and the Cosmological Constant,'' Nucl. Phys. {\bf B357} (1991) 534.}
\lref\SCH{J. H. Schwarz, Nucl. Phys. {\bf B226} (1983) 269. }
\lref\phil{P. Candelas, X. de la Ossa. P. Green and L. Parkes,
``A Pair of Calabi-Yau Manifolds as an Exactly Soluble Superconformal Field
Theory'' Nucl. Phys. {\bf B359} (1991) 21. }
\lref\agm{P. Aspinwall, B. Greene and D. Morrison, {\bf B416} (1994) 414.}
\lref\hbhst{S. Giddings, J. Harvey, J. Polchinski, S. Shenker and
A. Strominger, hep-th/9309152, Phys. Rev. {\bf D50} (1994) 6422. }
\lref\grn{B. Greene, ``Lectures on Quantum Geometry'', to appear.}
\lref\cpw{S. Coleman, J. Preskill and F. Wilczek, Phys. Rev. Lett. {\bf 67},
(1991) 1975; Nucl.Phys. {\bf B378} (1992) 175.}
%
%

\Title{\vbox{\baselineskip12pt
\hbox{hep-th/9504090}}}
{\vbox{\centerline{\bf{MASSLESS BLACK HOLES AND CONIFOLDS IN STRING THEORY} }}}
{
\baselineskip=12pt
\centerline{ Andrew Strominger }
\bigskip
\centerline{\sl Department of Physics}
\centerline{\sl University of California}
\centerline{\sl Santa Barbara, CA 93206-9530}
\centerline{ \it  andy@denali.physics.ucsb.edu }

\bigskip
\centerline{\bf Abstract}
Low-energy effective field theories arising from  Calabi-Yau string
compactifications are generically inconsistent or ill-defined at the classical
level
because of conifold singularities in the moduli space. It is shown, given
a plausible assumption on the degeneracies of black hole states,
that
for type II theories
this inconsistency can be cured by nonperturbative quantum effects: the
singularities are resolved by the appearance of
massless Ramond-Ramond black holes. The Wilsonian effective action
including these
light black holes is smooth near the conifold, and the singularity is
reproduced when they are integrated out. In order for a quantum effect to
cure a classical inconsistency, it can not be suppressed
by the usual string coupling $g_s$. It is shown how the required $g_s$
dependence arises as a result of
the peculiar couplings of Ramond-Ramond gauge fields to the dilaton.}
\Date{}
%

\newsec{Introduction}

In the last several years there has been spectacular progress in our
understanding of Calabi-Yau compactifications of string theory.
A fascinating and ubiquitous phenomenon, beautifully
exemplified in \phil, is the occurrence of conifold singularities
in the moduli space of classical string vacua. These are typically
real codimension two surfaces in the moduli space at which a dimension two or
three submanifold
of the Calabi-Yau space shrinks to zero size, and the curvature of the
moduli space metric blows up. They arise in virtually every
Calabi-Yau string compactification.

The mathematics of these singularities is well understood
\refs{\conif,\phxec,\phil,\phxe, \agm, \grn}, but the physics is not.
Consider a
time-dependent modulus field in a cosmological setting
which is slowly rolling around in a generic fashion.  Eventually it will
run in to
a conifold singularity. Low-energy effective field theory -- as
described by a
a sigma model whose target is the moduli space --  is then inadequate for
continuing the evolution because the sigma model equations of motion
become singular.  Our current understanding of strings on
Calabi-Yau spaces must therefore be regarded as seriously incomplete.

There are several reasons why it is important to understand the
physical behavior
of string theory near conifolds. One of these is that, in our efforts to
connect
string theory with reality, it is important to find generic features of
string
compactifications which might lead to model-independent predictions.
Conifold singularities are certainly one of the most significant such
features. Secondly, under some weak assumptions,
a generic superpotential will have local (or possibly global)
minima at the conifold singularities. It is thus quite possible that
the theory is driven to the conifold after supersymmetry breaking.
Since this point is outside the main  emphasis of the present paper its
demonstration is given in a brief appendix.

In this paper we endeavor to explain the physics of conifolds.
Specifically we show that, in the context of a type II string\foot{The
proper interpretation of conifolds for heterotic strings remains a mystery.},
the breakdown of low-energy
effective field theory may arise from integrating out a field which
becomes massless
at the singularity.  The quanta of this field are extremally-charged
Ramond-Ramond (RR) black holes! The conifold singularity can be
resolved by including this field in the low-energy effective action.

The consistency of our picture requires that there is one and only one
supermultiplet which becomes massless at the singularity. Ideally one would
like to derive this from a semiclassical analysis of the extremal black
holes. Unfortunately the counting of black hole states is a tricky problem,
so the best we can do at present
is give some plausibility arguments in section 4.2.
The consistency of our overall picture is of course evidence for
our assumption, but it is not ruled out that there is a completely
different resolution
of the conifold singularities with a different numbers of
massless states. Indeed, evidence for a picture quite different than that
described herein was given in \conif.

At first it may seem surprising that classical black holes can become
massless. However this phenomenon has an appealing explanation from
a ten-dimensional perspective.  The IIA (IIB) theory has extremal black
twobrane
(threebrane)
solutions \bhole\ whose mass is proportional their area.  After Calabi-Yau
compactification these may wrap around minimal two (three) surfaces
and appear as four-dimensional black holes. As the area of the surfaces
around which they wrap goes to zero, the corresponding
black holes become massless.

The structure of the low-energy effective field theory described here is
very similar to that found in \swit\ for $N=2$ Yang-Mills. This latter case
involves
conifold singularities which are resolved by the inclusion of massless
BPS magnetic monopoles, rather than extremal
black holes. A characteristic feature of both examples
are monodromies which describe how the charges of a state transported around
the singularity
are shifted by an $SP(2r;Z)$ transformation, where $r$ depends on the dimension
of the moduli space.

A puzzling difference between the two cases is as follows.
In the Yang-Mills case
the form of the singularity can be derived from a
one-loop computation in the light monopole
effective field theory, or equivalently as an instanton effect
in the original theory.
In our case it is possible to derive the form of the singularity by a
one-loop computation in the effective field theory with light black holes,
but we do not understand the analog of the instanton computation.
As discussed in section 4.3, such an instanton computation may
make sense in the context of an as-yet-unknown
dual formulation of string theory in
which fundamental strings are solitons.

It has long been speculated that black holes should be treated as
elementary particles.
In the present paper we argue that string theory requires this -
in the sense that extremal black holes contribute to virtual
quantum loops in a manner determined by their mass and charge -
in order to have a consistent resolution of conifold singularities.

In addition to conifold singularities, the moduli space of string vacua
can also contain orbifold-like singularities
at which low-energy field theory is in
danger of breaking down. It has been understood for some time
(see for example \dhs ) that such singularities can be resolved by the
appearance of extra light vector bosons associated with the singularity.
However for example in $K3$
compactification of a type II theory there are no such massless
states in string perturbation theory. It was conjectured in
\refs{\town,\wit,\ferr,\sentwo} and verified by construction in \jhasl\
that such states exist non-perturbatively with the correct degeneracies
as RR solitons.
The fact that the solitons have spin one as required for a gauge boson is
related to the $N=4$ supersymmetry. The resolution we propose here of conifold
singularities is similar in spirit. The difference in
the present case is that there is only $N=2$ supersymmetry, and the solitons
have
maximum spin one-half. They therefore produce conifold
rather than orbifold singularities when they become massless.

A striking feature of both the $N=4$ and $N=2$ cases is that consistency
of the low energy theory is rescued by the intervention of nonperturbative
quantum effects which are independent of the usual string coupling $g_s$
and are thus even larger than the $e^{-1/g_s}$ effects discussed in
\shenker. However the nature of these effects is quite unusual and
very restricted in form. We shall see in 4.1
that these large effects are possible due to the peculiarities of
RR gauge fields. Related observations were made in \wit.

This paper is organized as follows. Section 2 briefly reviews some
relevant aspects
of conifolds and special geometry. Section 3 describes the black holes
and their properties. In section 4 it is explained how they resolve the
conifold singularity, and we conclude in section 5. The appendix discusses
the behavior of a generic superpotential near a conifold.

\newsec {Review of Conifolds and Special Geometry}

This section contains a lightning review of some relevant aspects of
conifolds \phil\ and special geometry \refs{\dlwp, \ascmp}. More complete
discussion and references can be found in \phil\ - \grn.

A Calabi-Yau space $X$ has $b_3$ topologically non-trivial
three-surfaces.
Poincare duality implies the existence of a fixed integral basis
$A_I,~ B^J~I,J=1,..\half b_3$
of surfaces with intersections
\eqn\ints{A_I \cap B^J=-B^J\cap A_I
={\delta_I}^J,~~ A_I\cap A_J=B^I\cap B^J=0.}
This basis is unique up to $Sp(b_3;Z)$ transformations, under which
$(A_I,~ B^J)$ transforms as a vector and the intersection matrix \ints\
is preserved.
A choice of complex structure on $X$
is characterized by the  $b_3$ periods of the holomorphic three-form
$\Omega$
\eqn\perds{\eqalign{F_I&=\int_{A_I}\Omega,\cr Z^J&=\int_{B^J}\Omega,\cr}}
with respect to the fixed basis \ints. The periods $Z^I$ (or
alternately $F_I$) can be used as projective coordinates on the moduli
space $\cal M$ of complex structures on $X$. They are projective because
complex rescaling of $\Omega$ does not correspond to a change in the
complex structure. $\cal M$ is a special
Kahler manifold, and the periods \perds\ are a projective section of an
$Sp(b_3;Z)$ vector bundle over $\cal M$ \refs{\dlwp, \ascmp}.

In general $\cal M$ may contain a complex codimension one submanifold,
at which one of the periods,
say $Z^1$, vanishes. When $Z^1=0$, the corresponding three-surface
$B^1$ degenerates to zero size, and the Calabi-Yau space $X$
is a singular ``conifold''. Since the surface $Z^1=0$ is complex codimension
one
in $\cal M$, it can be encircled by a closed loop. If $X$ is
transported about this loop there is no guarantee that the
basis \ints\ will return to itself. Indeed
there is a typical $Sp(b_3;Z)$ monodromy\foot{This is the mondromy found
for the quintic conifold in \phil\ and appears to be generic. However there are
more complicated types of conifolds with multiple degenerations and different
monodromies. Further analysis willl be required to see if these can be resolved
in a similar fashion.}
\eqn\mnd{\eqalign{Z^1&\rightarrow Z^1,\cr F_1&\rightarrow F_1+Z^1.\cr}}
This implies that near $Z^1=0$
\eqn\fzm{F_1(Z^1) \sim {\rm constant}+{1\over 2 \pi i}Z^1\ln Z^1,}
while all the other periods are smooth and nonzero (for a simple degeneration).

The existence of a singularity in the metric $\cal G$ on $\cal M$ follows
readily from equation \fzm\ and the formula for $\cal G$ \tian:
\eqn\metfr{{\cal G}_{I\bar J}=\p_I\p_{\bar J}{\cal K},}
where
\eqn\klr{{\cal K}=-\ln (iF_I\bar Z^I -iZ^I \bar F_I).}
Substituting \fzm\ for $F_I$ one finds that for $Z^1$ near zero the
metric diverges as
\eqn\csn{{\cal G}_{1\bar 1}\sim \ln (Z^1 \bar Z^1).}
It is easily checked that the distance to $Z^1=0$ as measured by
$\cal G$ is finite and that the scalar curvature diverges there \phil.

To specify a string vacuum one must also choose a complexified
Kahler class $J+iB$ on $X$. This leads to $b_2$ additional complex moduli.
Mirror symmetry exchanges these with the complex structure moduli, implying a
parallel description in terms of periods of the complexified Kahler form
which can be found in \phxe. Of course the classical moduli space is just the
complexified Kahler cone and has no
conifold singularities. However string theory instructs us to correct for
the effects of worldsheet instantons which wrap around topologically
non-trivial
two-surfaces. One then finds that the periods can vanish at points where the
instanton corrections are large. This corresponds to a conifold singularity at
which the
quantum-corrected area of
a two-surface is degenerating to zero. Note that these are quantum corrections
on the
worldsheet only, and that no string loop effects are included in this
calculation.

\newsec{Extremal Black $p$-branes}

In this section we will argue that type II string theories
compactified on a Calabi-Yau space contain
black holes which become massless near a conifold.

\subsec{The Type IIB Theory}

Ten-dimensional type IIB string theory \SCH\ contains a
self-dual five-form field strength obeying
\eqn\ffd{\eqalign{F&=*F,\cr dF&=0,\cr}}
when the other  antisymmetric tensor fields in the theory vanish. There
is an associated conserved charge
\eqn\fchrg{Q(\Sigma_5)=\int_{\Sigma_5}F}
for every homology class of five-surfaces $\Sigma_5$.
The charge $Q$ can be carried by a $3+1$ dimensional
extended object, or threebrane enclosed by $\Sigma_5$.
Extended black hole solutions carrying the charge
$Q$ are given by \bhole\
\eqn\trbh{\eqalign{ds^2&=-(1-r_+^4/r^4)(1-r_-^4/r^4)^{-1/2}dt^2\cr
                       &~~~+{dr^2 \over (1-r_+^4/r^4)(1-r_-^4/r^4)}\cr
                       &~~~+r^2d\Omega_5^2+(1-r_-^4/r^4)^{1/2}dx_idx^i,\cr
                     F&=Q(\epsilon_5+*\epsilon_5),\cr
                     \phi &=\phi_0,}}
where $\phi$ is the dilaton, $\int_{S^5}\epsilon_5=1$, $x^i, ~i=7,8,9$ is
a coordinate on the threebrane, and the inner and outer horizons $r_\pm$ are
related to $Q$ by $Q=2r_+^2r_-^2$. $r_+$ is a regular outer horizon and there
is a singularity at $r_-$. Classically there is a solution for every value of
$Q$, but we will assume that there is a quantization condition so that $Q$
takes the minimal value $Q=g_5$.

At the quantum level most of these solutions are unstable due to
Hawking radiation. They will decay to the extremal limit
at which
$r_+=r_-$ and the metric becomes
\eqn\trbhe{\eqalign{ds^2&=-(1-r_+^4/r^4)^{1/2}dt^2\cr
                       &~~~+{dr^2 \over (1-r_+^4/r^4)^2}\cr
                       &~~~+r^2d\Omega_5^2+(1-r_+^4/r^4)^{1/2}dx_idx^i,\cr} }
and the solution is supersymmetric \refs{\bhole,\duffthrb}.
A Bogolmonyi bound implies
\eqn\bbnd{{\rm Mass}=Q\times ({\rm 3-volume}).}

  Now consider compactification of the IIB theory from ten to four dimensions
on a Calabi-Yau space
$X$.  The four-dimensional theory will have $h_{21}$ $N=2$ vector multiplets
whose
scalars are the coordinates $Z^I$  on $\cal M$ with metric $\cal G$ of \metfr.
The
extremal threebrane can then wrap around one of the
three-surfaces $A_I,B^J$  in $X$.  Such configurations will appear as
ordinary localized solitons or black holes in four dimensions.
Quantiztion of RR charge implies
\eqn\fqnt{\eqalign{\int_{A_I\times S^2}F &=n_Ig_5, \cr
               \int_{B^J\times S^2}F &=m^Jg_5, \cr}}
where $S^2$ is a spatial two-sphere surrounding the black hole.
The integers $m^J$ and $n_I$ are four-dimensional electromagnetic
charges associated with the $h_{21}$ vector multiplets and
the graviphoton.
One expects that
the minimal energy configuration will saturate a BPS bound\foot{A
closely related class of black hole solutions saturating BPS bounds
appears in \master.}.
Up to an overall constant the unique Kahler and  $Sp(b_3;Z)$
invariant formula
for the BPS mass
is \ferr\
\eqn\mfrm{M=g_5 e^{{\cal K}/2}|m^IF_I-n_IZ^I|.}
Of special interest is a  threebrane with all charges equal to
zero except $n_1=1$.   Its mass is then
\eqn\mfrzm{M=g_5 e^{{\cal K}/2}|Z^1|,}
which vanishes at the conifold $Z^1=0$. The reason for this can be
heuristically
understood as follows.
Accordnig to \bbnd\ the mass is proportional the area
of the minimal surface around which the
threebrane wraps. Practically by definition a
conifold is a point at which this minimal area vanishes.

\subsec{The Type IIA Theory}

The type IIA theory has a four-form RR field strength $F$. The associated
charge is carried by twobrane solutions \bhole.  Calabi-Yau
compactification on $X^\prime$ leads to $b_2$ vector multiplets
whose scalars are
again the coordinates on a special Kahler manifold ${\cal M}^\prime$.
The twobranes
can wrap around minimal two-surfaces of ${\cal M}^\prime$ to produce
four-dimensional extremal black holes. These black holes will
then become massless at the conifold singularities of ${\cal M}^\prime$
where the area of the minimal surface (corrected by worldsheet instantons)
degenerates.
Since mirror symmetry exchanges IIB compactification on $X$
with
IIA compactification on the mirror of $X$ \refs{\nati,\serg}, it will also
exchange
threebranes with twobranes. Hence if $X^\prime$ is the mirror of $X$ the
analysis of the black hole solutions reduces to that considered above.

While the mathematical analysis is equivalent, the physical picture is
significantly different. In the IIA case the conifold singularity and the
massless states are a result of large instanton corrections. This will
figure in to the comparison of section 4.3 with  recent quantum results
\swit\ in d=4 supersymmetric gauge theories.

\newsec{Resolving the Conifold Singularity}

In this section we  explain, following a parallel discussion in \swit, how the
appearance of a massless black holes resolves the puzzle of conifold
singularities
mentioned in the introduction.
We wish to understand the four-dimensional, $N=2$ supersymmetric, low-energy
effective action resulting
from Calabi-Yau compactification of type IIB string theory.  At generic points
in
the moduli space, the massless fields will consist of the graviton multiplet,
$h_{21}$ vector multiplets and $b_2+1$ hypermultiplets. We shall focus on the
conifold singularities in the moduli space of the vector multiplet.

Near a conifold at $Z^1=0$, the BPS state with $n_1=1$ and all other
charges equal to zero has a mass which vanishes as in \mfrzm. This state
carries the
minimal charge with respect to the $U(1)$ gauge field which lies in a
supermultiplet with the moduli field $Z^1$, and is itself part of a
hypermultiplet \gihu. The low-energy
effective action involving only moduli fields must break down near $Z^1=0$
because this state becomes light and can be excited. It should be replaced by
an effective field theory in which the kinetic term of the $Z^1$ modulus does
not
diverge and there is an additional charged hypermultiplet. This is
basically identical\foot{A fascinating difference is that in \swit\
a complex torus with
degenerating cycles was introduced as a mathematical artifact to solve the
equations determining the moduli space. In our case  both the analog of the
torus -
the Calabi-Yau manifold - and the analog of the cycles are not introduced
just to solve the equations but are ``really there''. } to a situation
described in \swit.
In the low energy Wilsonian effective action
there are no logarithms in $F_1$. However the one loop
beta function obtains
a contribution from the light charged hypermultiplet. Since the
gauge coupling is a derivative of $F_1$, this implies a one
loop correction to $F_1$ after integrating
out the black hole hypermultiplet. This leads to
\eqn\flp{F_1(Z^1) \sim {\rm constant}+ {1\over 2 \pi i}Z^1\ln Z^1,}
in agreement with \fzm. Thus the conifold singularity
in the metric at $Z^1=0$ is just the usual type of singularity produced by
intergrating out massless charged fields.

\subsec{Classical vs. $g_s\rightarrow 0$ Limit}

There is something puzzling about the preceding comments. In \swit\
the singularity in the moduli space arises from
nonperturbative quantum effects and is resolved
by quantum loops of light monopoles. In our example the singularity arises
in the classical moduli space. Naively it would seem impossible to resolve
such a singularity by quantum loops of anything, since these should be
suppressed
as $g_s=e^{\phi}$ (where $\phi$ is the string dilaton) vanishes.

The surprising fact\foot{Observations closely related to the following
were made in
\refs{\town,\jhasl,\sentwo} and especially \wit.} is that non-perturbative
quantum effects
are not all
suppressed as $g_s\rightarrow 0$. Usually we think that in string theory
$\hbar $ appears in the combination $\hbar e^{2 \phi}$ because
a genus $g$ surface is always accompanied by a factor of
$ e^{2 (g-1)\phi}$. This suggests that as
$e^{\phi} \rightarrow 0$ quantum effects are turned off. This argument
has a loophole
in theories with RR fields. The classical
action
is most naturally written in the form
\eqn\rract{S \sim\int(e^{-2 \phi}R+F^2+...)}
where $F$ is a RR field strength. This seems to contradict the expectation from
string perturbation theory that the classical action should be proportional to
$e^{-2 \phi}$, but it is not a contradiction because \refs{\wbran,\wit} this
factor
could be reinstated in front of the second term by rescaling the potentials.
However this would lead to peculiar gauge transformation laws
and quantization conditions. There is no reason
why nonperturbative effects associated with \rract, which does not have a
uniform overall factor of $1/g_s^2$, should behave as
$e^{-1/g_s^2}$.

We wish to see that, after an appropriate rescaling,
loop effects of RR black holes contain no factors of
$g_s$. The fastest way to see this is to note that after rescaling
the metric by  $e^{-2 \phi}$, the four dimensional effective
action at generic points in the moduli space can be written in
the form\foot{These are the variables in which spacetime
supersymmetry is manifest.}
\eqn\rfact{S_4 \sim\int d^4x\sqrt{-g}(R+ ({\rm vector~multiplets})+
({\rm hypermultiplets})+...).}
The dilaton $\phi$ lives in a hypermultiplet. $N=2$ supersymmetry forbids
neutral couplings between vector multiplets and hypermultiplets\foot{This
implies a nonperturbative nonrenormalization theorem for the metric \metfr\
\bbs.} \dlwp.
Thus the first two terms in the action are simply unaware of the value of
$g_s$. The black holes which becomes massless at the conifold points
are soliton solutions involving only these two terms and are therefore also
unaware of the value of
$g_s$. In a low energy effective field theory they will be represented by a
hypermultiplet with a $g_s$-independent action. Therefore quantum loops of
these solitons do not involve $g_s$.

In conclusion, the $g_s \rightarrow 0$ limit of the low-energy effective action
of the full quantum theory is very different than the
low-energy effective action at string tree-level.  The latter is inconsistent
because of conifold singularities. The inconsistency may be resolved by the
inclusion of non-perturbative quantum effects.

\subsec{Degeneracies of States}

A second puzzle concerns the number of states which become
light near a conifold. The mass of a state with $n_1=2$ also
vanishes. If this is interpreted as a charge-two, single-particle state,
there would be an extra contribution to the beta function, ruining the
agreement between \fzm\ and \flp. We therefore assume that this does not
correspond to a single particle state. Rather it is a two-particle state
consisting of two charge-one particles.

The counting of extremal black hole states is a
long-standing unsolved problem in quantum gravity. We do not
know how to derive our assumption from first principles (except from the
requirement of a consistent low-energy theory), but it can be supported by
consideration of some other examples.  A first example is the
`t Hooft-Polyakov monopole. The charge-two solution is just a point in the
moduli space of two charge-one solutions, and there is no separate charge-two
monopole.
However black hole moduli spaces are less well defined near the coincident
point and the
question is not so easy to answer. In a second example \jhasl, minimally
charged
string solitons were
constructed as extremal black holes in a $K3$ compactification of type IIA
string theory. Closed loops of these solitons had just the right masses and
degeneracies to
provide the enhanced gauge symmetries at the points in moduli space where the
$K3$ surface degenerates. However the soliton string solution can carry any
axion charge. If the charge-two string had been viewed as distinct from two
charge-one strings, a consistent picture would not have emerged. A related
example is a fundamental string solution\refs{\DAHA,\DGHR}, which has just the
right
charge/mass ratio to be viewed as an extremal black hole \bhole. If we treated
the
multiply-charged solutions as distinct objects, there would
be an infinite number of gravitons!

Other examples however seem to go against our assumption. In \wit,
the relation between type IIA string theory and $d=11$ supergravity seems to
require an infinite tower of distinct multiply-charged states.
Also the spectrum of Dabohlkar-Harvey winding states \refs{\DAHA,\DGHR}
of heterotic string
theory at charge two contains both a two-string state
and a doubly-wound, one-string ``bound state at threshold''.
The higher $n_1$ states discussed here might be viewed as a similar type of
winding state.

To summarize, consistency of the low energy theory seems to require that
ther are no multiply-charged single-particle states. It would be
interesting
to verify (or disprove!) this from some other starting point,
such as possibly a semiclassical analysis of the solutions and their moduli
spaces.
\subsec{Quantum Duality}

A further puzzle concerns the fact that the conifold singularities
discussed herein
arise in what is
usually referred to as classical string theory, as opposed to the
conifold singularities
of $N=2$, $d=4$ Yang-Mills \swit, which are not present classically
but arise from quantum instanton corrections.  This seems to be a radical
difference. However there have been
conjectures \refs{\DGHR,\DAHA, \hetsol,\wbran,\duffcq, \duffss,\wit } that
there
exists a dual formulation of string theory in which fundamental strings
arise as solitons.  In any such dual formulation of string theory, a
worldsheet instanton of the ordinary formulation of string theory will be
a spacetime instanton leading to nonperturbative quantum effects.  In type IIA
string theory, the conifold singularities are
produced by large effects from worldsheet instantons.
In a dual formulation of the IIA string, the conifold singularities
would then be interpreted as a nonperturbative instanton effect,
exactly as in the $N=2$ Yang-Mills
case.  Thus the idea of a dual formulation of string theory would
dovetail nicely with the observations of this paper.

Related phenomena were discussed in \hbhst.  In \cpw, an Abelian Higgs model,
which has soliton strings, was coupled to gravity. Instantons were
found corresponding to a soliton string worldsheet wrapping around a black hole
horizon.  These instantons give nonperturbative corrections for
example to the
Hawking temperature. The strength of these effects goes as
\eqn\scpw{e^{-TA/\hbar },}
where $T$ is the soliton string tension and  $A$ is the area of the
horizon.  In \hbhst\ the same effect was considered in the context of
fundamental string
theory.  In this case the instantons are fundamental string worldsheets
wrapping around a black hole horizon, {\it i.e.} worldsheet instantons.
The strength of the effects goes as
\eqn\shbh{e^{-A/2\pi \apm }.}
This expression does not involve $\hbar$ and so is viewed as a classical
effect.
However \shbh\ reduces to \scpw\ when reexpressed in terms of the string
tension
$T=\hbar /2\pi \apm$.  Similarly the worldsheet instantons which produce
type IIA
conifold singularities become ordinary spacetime instantons when the
worldsheet is
itself a soliton.

\newsec{Conclusions}

In closing we mention that our analysis has brought to light several
interesting
features of string theory whose significance goes beyond the example
discussed
here:

\noindent 1. Nonperturbative quantum effects of a certain restricted type
may occur in string theory
which are
not suppressed as $g_s \rightarrow 0$.

\noindent 2. Classical string theory is inconsistent without the inclusion
of
such effects.

\noindent 3. The quantum consistency of string theory requires
that extremal black holes be treated as elementary quanta.

\centerline{\bf Acknowledgements}
I am grateful to K. Becker, M. Becker, P. Candelas, B. Greene, J.
Harvey, D. Lowe, E. Martinec, J. Polchinski, N. Seiberg, L. Thorlacius and
M. Srednicki for useful discussions.
This work was supported in part by DOE grant DOE-91ER40618.

\appendix{A}{Superpotentials}

In this appendix we consider the behavior of a superpotential near a
conifold. We consider an
$N=1$ theory with  moduli fields parameterizing a space with a conifold
singularity.
This arises in Calabi-Yau compactification of heterotic string theory.

The superpotential $W$ is a section of a line bundle $L$ whose curvature
is the
Kahler form \bwit. $L$ has no monodromy about a conifold singularity and
$\cal K$ is finite there.  A generic superpotential can therefore be
expanded about
a conifold singularity at $Z^1=0$ as
\eqn\wexp{W=w_0+w_1Z^1+....}
The scalar potential is related to $W$ by
\eqn\vw{V=e^{\cal K}{\cal G}^{I \bar J}D_I W D_{\bar J}\bar W-3e^{\cal K}
W \bar W.}
Near the conifold the inverse metric behaves as
\eqn\gbhv{{\cal G}^{1 \bar 1} \sim {1 \over {\rm ln}(Z^1\bar Z^1)}+... .}
$V$ will then behave as
\eqn\vbhv{V\sim V_0 + {w_1^2\over {\rm ln}(Z^1\bar Z^1)}+....,}
which has a sharp minimum at $Z^1=0$.  Thus it is plausible that
a superpotential could have local minima at conifold singularities.

Of course the $N=1$ supersymmetry present in heterotic string theory
allows perturbative corrections to $\cal G$. These could qualitatively
alter the geometry near the conifold, so these observations
should be regarded as merely suggestive.

\listrefs

\end